\documentclass[lettersize,journal]{IEEEtran}
\usepackage{amsmath,amsfonts}
\usepackage{algorithmic}
\usepackage{algorithm}
\usepackage{array}
\usepackage[caption=false,font=normalsize,labelfont=sf,textfont=sf]{subfig}
\usepackage{textcomp}
\usepackage{stfloats}
\usepackage{url}
\usepackage{verbatim}
\usepackage{graphicx}
\usepackage{cite}
\usepackage{multirow}
\usepackage{subfig}
\begin{document}

\title{YMIR: A new Benchmark Dataset and Model for Arabic Yemeni Music Genre Classification Using Convolutional Neural Networks}

\author{Moeen~AL-Makhlafi,
	Eiad~Almekhlafi,
	Abdulrahman A. AlKannad,
	Nawaf Q. Othman,
	Ahmed Mohammed,
	and Saher Qaid
	
	\IEEEcompsocitemizethanks{ \IEEEcompsocthanksitem M. AL-Makhlafi, A. AlKannad, N. Othman A. Mohammed and S. Qaid are School of Artificial Intelligence, Xidian University, Xi’an, China.
		\IEEEcompsocthanksitem E. Almekhlafi is with Department of Information Science and Technology, Northwest University, Xi’an 710127, China.}
	\thanks{Manuscript received September 09, 2023; revised August 26, 2015.}}

\markboth{Journal of \LaTeX\ Class Files,~Vol.~14, No.~8, August~2021}%
{Shell \MakeLowercase{\textit{et al.}}: A Sample Article Using IEEEtran.cls for IEEE Journals}


\maketitle

\begin{abstract}
Automatic music genre classification is a major task in music information retrieval; however, most current benchmarks and models have been developed primarily for Western music, leaving culturally specific traditions underrepresented. In this paper, we introduce the Yemeni Music Information Retrieval (YMIR) dataset, which contains 1,475 carefully selected audio clips covering five traditional Yemeni genres: Sana'ani, Hadhrami, Lahji, Tihami, and Adeni. The dataset was labeled by five Yemeni music experts following a clear and structured protocol, resulting in strong inter-annotator agreement (Fleiss' $\kappa$= 0.85). We also propose the Yemeni Music Classification Model (YMCM), a convolutional neural network (CNN)-based system designed to classify music genres from time-frequency features. Using a consistent preprocessing pipeline, we perform a systematic comparison across six experimental groups and five different architectures, resulting in a total of 30 experiments. Specifically, we evaluate several feature representations, Mel-spectrograms, Chroma, FilterBank, and MFCCs with 13, 20, and 40 coefficients, and benchmark YMCM against standard models (AlexNet, VGG16, MobileNet, and a baseline CNN) under the same experimental conditions. The experimental findings reveal that YMCM is the most effective, achieving the highest accuracy of 98.8 \% with Mel-spectrogram features. The experimental findings not only reveal that YMCM is the most effective but also provide practical insights into the relationship between feature representation and model capacity. The findings put YMIR at a good benchmark and YMCM at a strong baseline for classifying the genre of Yemeni music.
\end{abstract}


\section{Introduction}
Music is culturally and socially significant, taking on diverse forms and styles around the world. The wide range of personal music tastes makes precise music classification a core task for creating personalized recommendations \cite{elbir2020music}. Since song titles alone are not enough tools for classification, musical genre has proven to be one of the most reliable criteria for this task \cite{aucouturier2003representing}. Although genre is often assigned by hand, it remains essential for matching recommendations to listeners' tastes. Major services like Spotify and SoundCloud use genre to classify music, helping them engage users with personalized content \cite{elbir2018music}.

The exponential growth of multimedia content across a wide range of digital platforms has substantially exacerbated the difficulties involved in effectively indexing, browsing, and retrieving music files. In this context, automatic music genre classification has become increasingly important for organizing large audio collections. It involves analyzing different musical characteristics, such as timbre, instrumentation, and lyrics \cite{serwach2016ga}. Although classifying music genres is difficult because music is complex and varied, each genre usually has clear patterns. Good classification becomes possible when we properly model how the different features relate to each other.

Several techniques for automatic music classification have been introduced \cite{tzanetakis2002musical, wold1996content}, but most were developed and evaluated only on well-known Western music datasets. However, there is a great lack of research on the classification of Arabic music, especially when it comes to Yemeni music genres.

Yemeni music ranks among the oldest and most culturally rich musical traditions in the Arab world. It comes from centuries of oral poetry, religious practices, and social traditions. This music shows Yemen's rich and varied history, geography, and ethnic groups. Yemeni music can more accurately be seen as one of the main roots of Arab music \cite{lambert2020historical, lavin2020colonial, ahmed2007yemeni, altwaiji2024folk, unesco2003sanani}.

A major challenge in classifying Arabic music, especially Yemeni music, is the serious shortage of available training data. In this paper, we have tackled this issue by generating the Yemeni Music Information Retrieval (YMIR) dataset, which encompasses data for the five primary genres: Sana'ani, Hadhrami, Tihami, Lahji, and Adeni. Additionally, we have proposed the Yemeni Music Classification Model (YMCM), a genre classification system built upon the widely recognized Convolutional Neural Network (CNN) architecture. We subsequently carried out six main experimental groups, with each group including five sub-experiments. In the six main experiments, we used different time-frequency feature extraction methods: Mel-frequency Cepstral Coefficients (MFCCs) with 13, 20, and 40 coefficients (MFCC13, MFCC20, MFCC40), Mel spectrograms, Chroma features, and FilterBank features, all extracted from the YMIR dataset. For each feature type, we fed the extracted features separately into five different convolutional neural network models: our proposed YMCM, AlexNet, a standard CNN, VGG16, and MobileNet. This setup produced a total of 30 distinct experiments.

The main contributions of this work are summarized as follows:

\begin{itemize}
	
	\item We release YMIR, the first expert-annotated dataset for Yemeni music genre classification, comprising 1,475 audio clips across five traditional genres (Sana’ani, Hadhrami, Lahji, Tihami, and Adeni). The dataset was labeled by five Yemeni music experts using a structured protocol, achieving strong inter-annotator agreement (Fleiss' $\kappa$=0.85).
	
	\item We propose the Yemeni Music Classification Model (YMCM), a CNN-based architecture with five convolutional layers designed for genre classification from time–frequency representations.
	
	\item  We provide a consistent preprocessing and segmentation workflow and adopt stratified training/testing splits to support fair and repeatable evaluation.
	
	\item We systematically compare six of the feature extraction techniques: Mel-spectrograms, MFCCs with 13, 20, and 40 coefficients, Chroma, and FilterBank features, and quantify their effect on classification accuracy and stability. The results indicate that Mel-spectrogram features realize the highest accuracy on the YMIR dataset when used with the proposed YMCM model, outperforming the other feature extraction techniques.
	
	\item We benchmark YMCM against established architectures (AlexNet, VGG16, MobileNet, and a baseline CNN) under identical settings, resulting in 30 experiments (six feature sets × five models). YMCM achieves the best overall consistent performance, reaching 98.83\% accuracy.
	
\end{itemize}

\section{Previous research on music genre classification}

Music genre classification has been a well-explored area in Music Information Retrieval (MIR), with several techniques employing deep learning models for enhanced performance. In their seminal paper, Tzanetakis and Cook \cite{tzanetakis2002musical} define music genres as categorical labels that classify musical pieces based on elements such as instrumentation, rhythmic structure, and harmonic content. The authors identified three key features for analyzing musical content: timbral texture, rhythm, and pitch, particularly in the context of Western music styles like classical, jazz, pop, and rock. This foundational work has significantly influenced subsequent research in genre classification. Both entire recordings and homogeneous segments within those recordings were utilized, achieving a classification accuracy of 61\% across ten genres, using statistical pattern recognition classifiers. These results were closely aligned with those found in human genre classification studies.

A notable contribution to this field is the work by Han Ding (2024), who proposed a novel hybrid model combining Residual Networks (ResNet) with Bidirectional Gated Recurrent Units (Bi-GRU) for music genre classification. This approach leverages visual spectrograms as input, enabling the model to benefit from both the spatial feature extraction capabilities of ResNet and the temporal modeling capabilities of Bi-GRU. This method achieved promising results, showing the potential for deep learning models to significantly improve genre classification accuracy by capturing intricate patterns within the audio data \cite{ding2024genre}.

In a similar vein, Oguike and Primus (2025) introduced a multimodal classification system for Sotho-Tswana musical videos, incorporating audio, text (lyrics), and visual modalities. By using deep learning models for each modality and applying a decision-level fusion technique, their system demonstrated superior performance compared to unimodal models that rely solely on audio or lyrics \cite{oguike2025multimodal}.

Another innovative approach by Shen and Xiao (2024) applies Functional Data Analysis (FDA) to represent music signals as continuous functions, capturing both temporal and harmonic properties of music. This method, combined with Adaptive Fourier Decomposition (AFD), was tested on the GTZAN and FMA datasets, yielding significant improvements in classification accuracy over the traditional method \cite{shen2024music}.

Ahmed et al. (2024) explored the use of advanced deep learning models for music genre classification, comparing the effectiveness of various models such as CNNs, LSTMs, and SVMs. Their study highlights the superiority of CNNs in capturing complex spectrogram patterns, achieving high classification accuracy on the GTZAN and ISMIR2004 datasets \cite{ahmed2024musical}.

Beyond classical CNN baselines, recent studies have increasingly emphasized three practical levers for improving genre recognition: (1) augmentation and training-time regularization to mitigate limited dataset size, (2) systematic tuning of network hyperparameters, and (3) attention-based modeling for richer time-frequency context. For example, data augmentation coupled with deep architectures has been reported to substantially improve Mel-spectrogram-based classification on standard benchmarks \cite{ba2025augmentation}. Complementarily, automated configuration and hyperparameter optimization strategies have been explored to stabilize CNN performance across alternative spectral representations such as MFCC and STFT \cite{li2024heliyon}. In a separate line of work, hybrid Transformer designs have been proposed to strengthen feature extraction from Mel-spectrograms by combining convolutional locality with global self-attention and channel-wise emphasis \cite{wu2025vit}. These trends collectively suggest that, for fair comparison on low-resource regional corpora, it is essential to control preprocessing, segmentation, and evaluation splits while benchmarking both feature representations and model capacity under identical experimental settings. This motivates the development of expert-labeled regional datasets and controlled baselines that make cross-model and cross-feature comparisons reproducible, which is the objective of the YMIR dataset and the proposed YMCM evaluation protocol.

\section{Design of YMIR}

This dataset is the first publicly accessible compilation of Arabic songs, with a particular focus on Yemeni music genres, offering a robust foundation for research into regional musical styles. The Yemeni Music Information Retrieval dataset encompasses data from the five main genres: Sana’ani, Hadhrami, Tihami, Lahji, and Adeni. Each musical scale also exhibits unique stylistic traits, making its identification closely related to genre classification in other types of music. Classification accuracy was assessed through inter-annotator agreement. Ultimately, the music recordings were labeled and organized to create the YMIR dataset.

\subsection{Data Collection}

\begin{figure}
	\centering
	\includegraphics[width=90mm]{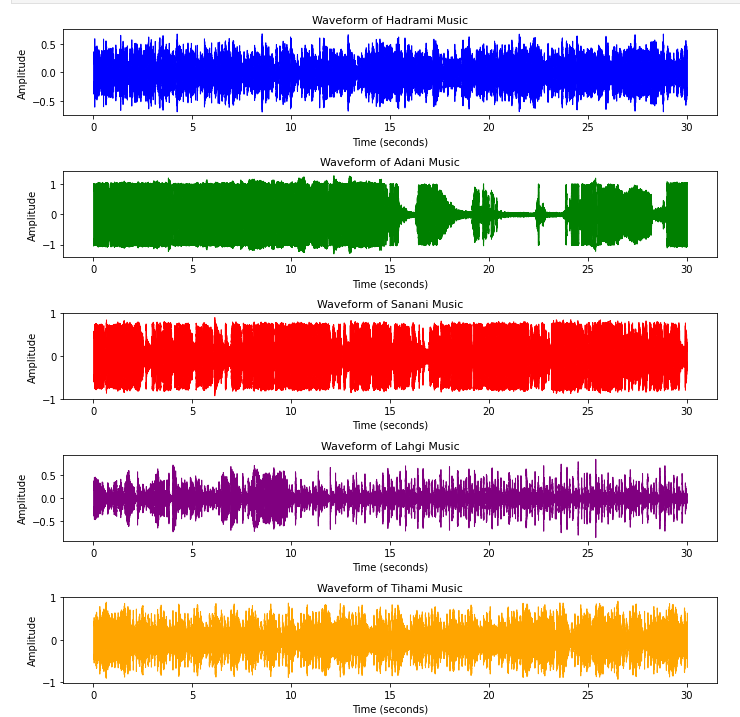}
	\caption{Waveform samples of the five YMIR genre classification dataset.}
	\label{Waveform}
\end{figure}

The dataset comprises 1,475 audio clips across five traditional Yemeni music genres: Sana'ani, Hadhrami, Lahji, Tihami, and Adeni. Each genre contains 295 audio files, with all clips standardized to a duration of 30 seconds. This standardization ensures a balanced representation across genres while also safeguarding the copyright of the original works.

The YMIR dataset was compiled from online sources, including platforms such as YouTube. The audio files are provided in WAV format, with a sample rate of The YMIR dataset was compiled from online sources, including platforms such as YouTube. The audio files are provided in WAV format, with a sample rate of 48 kHz and a bit rate of 360 kbps. The dataset is organized into multiple folders, each corresponding to one of the five song genres. Accompanying metadata files are included, detailing each clip's title, artist, genre, and file name. Fig. ~\ref{Waveform} shows the waveform representation of sample audio signals from the YMIR dataset.

\begin{figure}[!b]
	\centering
	\includegraphics[width=80mm]{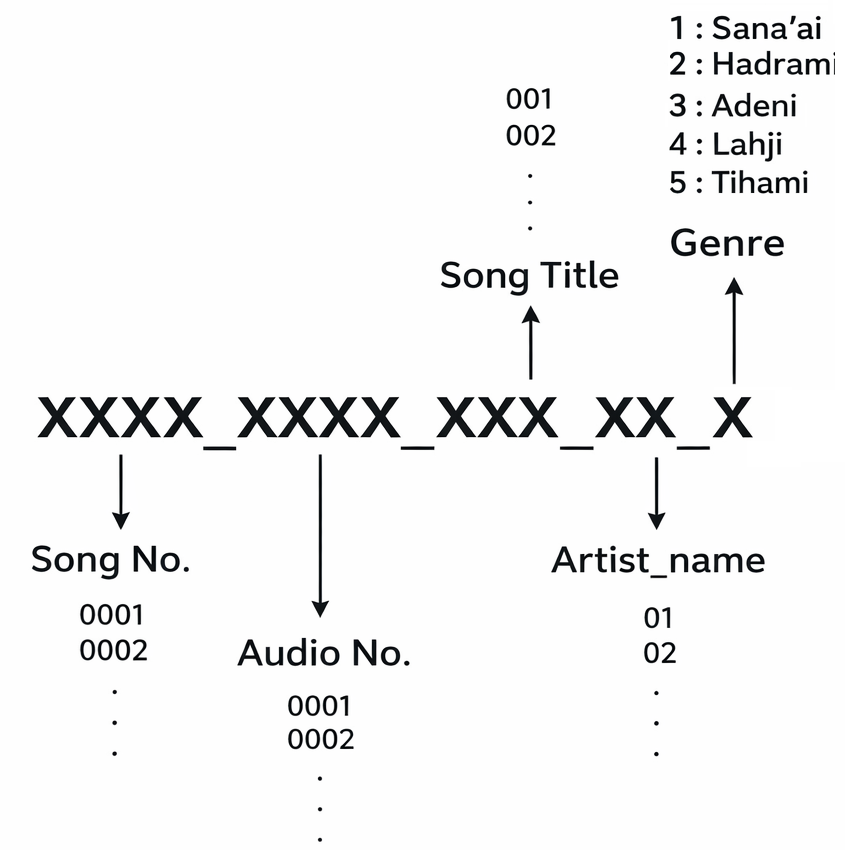}
	\caption{Data Labeling structure}
	\label{Labeling}
\end{figure}

\subsection{Data Labeling}

Experts in Yemeni music performed manual labeling to assign each audio track to its appropriate genre. Each track was labeled with a five-digit string, separated by underscores, representing its genre and position within the dataset. The string consists of the following elements: Song number, Sample number, Song title, Artist name, and Genre (as illustrated in Fig. ~\ref{Labeling}). This manual labeling process ensures the accuracy of classification tasks and facilitates the structured organization of the dataset.


\subsection{Judgments}

Five independent annotators, each possessing extensive expertise in Yemeni music genres, contributed to the labeling process. The annotators were provided with comprehensive guidelines to ensure consistency and accuracy in their assessments. They were instructed to attentively listen to each audio clip and assign the most appropriate genre from the five core categories: Sana'ani, Hadhrami, Lahji, Tihami, and Adeni.

Each annotator independently listened to all recordings and either assigned a genre or rejected the track if it did not clearly fit one of the five categories. In cases of disagreement, a consensus was reached through discussion, consultation, or majority voting. If three or more annotators agreed on a genre, the label was accepted for inclusion in the YMIR dataset; if not, it was rejected. For particularly ambiguous cases, annotators were advised to leave the track unannotated to avoid mislabeling. This process ensured consistency, accuracy, and reliability in the final labels.

To evaluate the reliability of the dataset, we calculated Cohen's Kappa score \cite{randolph2005free}, a statistical measure that assesses inter-rater agreement for categorical classifications, given that five judges were involved in the labeling process.

\begin{equation}
	\kappa = \frac{\bar{P}_0 - \bar{P}_e}{1 - \bar{P}_e}
\end{equation}

The factor \( 1 - \bar{P}_e \) how much the annotators agree with each other, beyond random guessing, while it \( \bar{P}_0 - \bar{P}_e \) indicates how much real agreement there is, above and beyond random chance. A value of \( \kappa = 1 \) signifies perfect agreement among all raters. Measuring agreement among the annotators for our dataset, as measured by Fleiss' kappa, yielded a score of 0.85, indicating a high level of agreement among the five raters.

\section{Methodologies}

\begin{figure*}
	\centering
	\includegraphics[width=130mm]{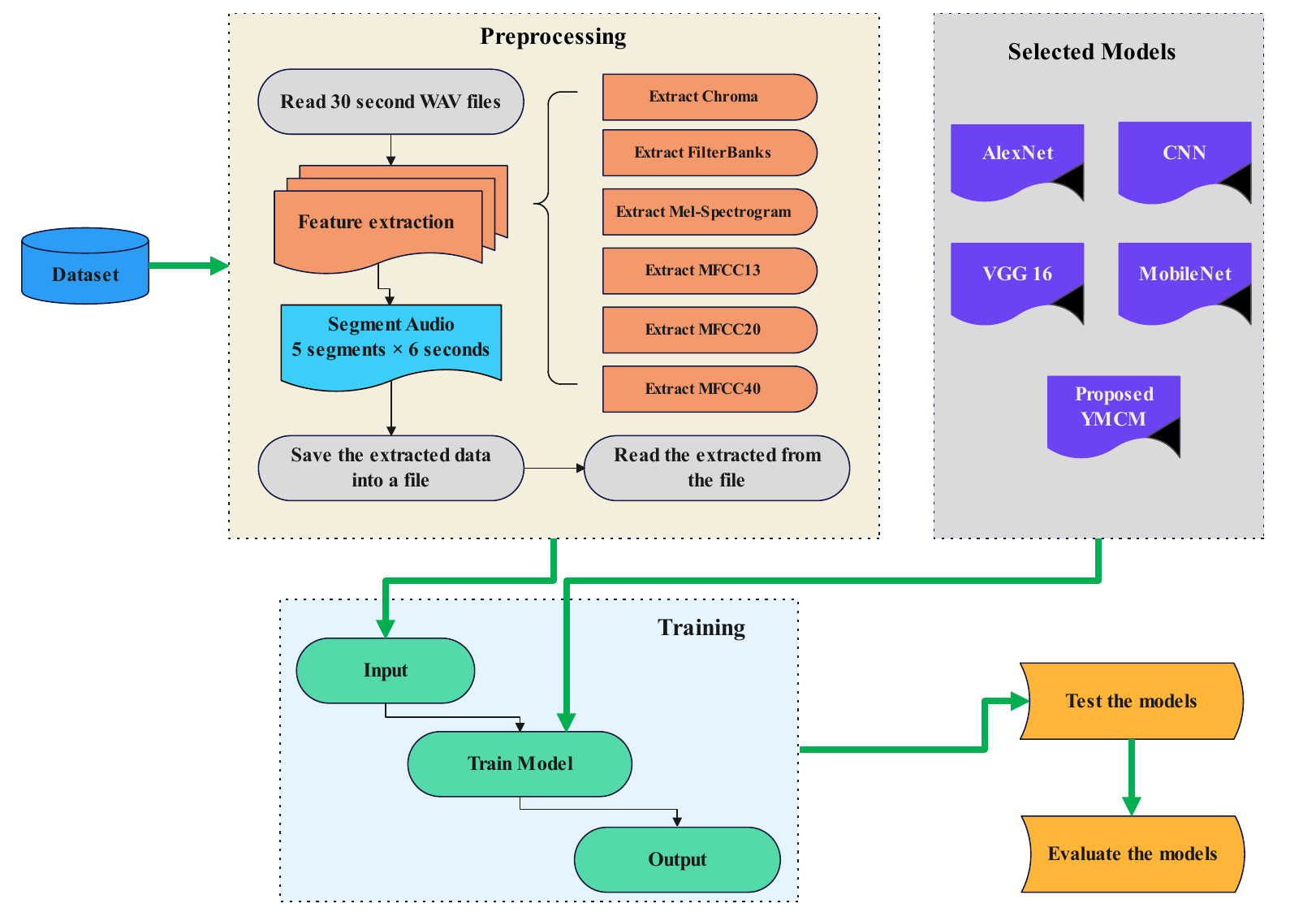}
	\caption{Workflow diagram of the proposed framework}
	\label{workflow}
\end{figure*}

The methodology adopted in this study is described in this section, covering key components including data preprocessing steps, feature extraction techniques, and the foundational deep learning architectures serving as baseline models for the proposed music genre classification system. An overview of the study's complete workflow is illustrated in Fig.~\ref{workflow}.

\subsection{Data preprocessing}

\begin{figure}
	\centering
	\includegraphics[width=75mm]{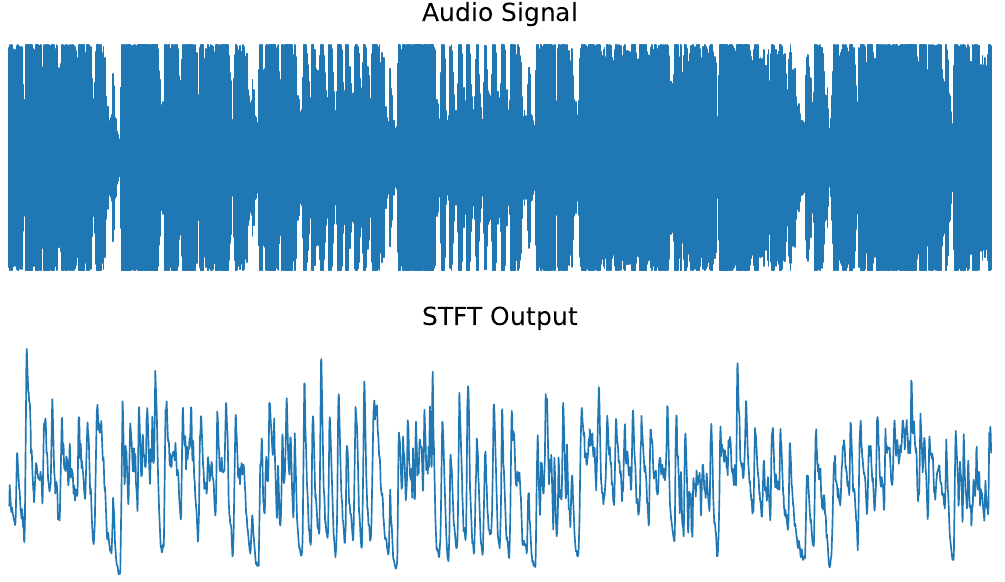}
	\caption{WAV-format audio data Short-Time Fourier Transform Extraction.}
	\label{STFT1}
\end{figure}

In this study, all audio signals were processed using a unified preprocessing pipeline to ensure consistency across the dataset. Each audio file was resampled to 22,050 Hz and truncated to a fixed duration of 30 seconds. To handle the non-stationary characteristics of musical signals, a Short-Time Fourier Transform (STFT) is applied to obtain a time–frequency representation. The resulting power spectrogram is computed once and subsequently reused to extract features, ensuring consistent spectral alignment across all representations while reducing redundant computations, as illustrated in Fig. ~\ref{STFT1}. To increase the number of training samples and capture temporal variations within each recording, every audio file was divided into five equal-length segments of 6 seconds each. After segmentation and preprocessing, the final dataset contained 7,258 samples, which were split into training and test sets using an 80:20 stratified split. This yielded 5,806 training and 1,452 testing samples, preserving the class distribution across both sets.

\subsection{Feature extraction}

The main purpose of feature extraction is to create a sequence of feature vectors that offer a compact yet informative representation of the input audio signal. In music classification, feature extraction is a crucial step because the quality and relevance of the extracted features directly determine how accurate and effective the classification will be. Paying close attention to this step is essential, as it directly affects how well the following classification algorithms perform.

As discussed earlier, music genre classification involves selecting suitable audio features and designing an effective classification model. Earlier research on music genre classification has mainly relied on four key feature types: Mel-spectrograms \cite{chillara2019music,dhall2021music, tang2020combining, ghildiyal2020music}, FilterBanks \cite{fayek2016speech}, Chroma \cite{zhang2014verification, singh2021deep, pulipati2021music},  and Mel Frequency Cepstral Coefficients (MFCC) \cite{kizrak2014classification, thiruvengatanadhan2018music, mandal2020automatic, sharma2021classification}. In light of these approaches, we incorporated all four feature types in our experiments to evaluate their performance and determine which would yield the best results for the YMIR dataset.

\subsubsection{Chroma}

Chroma features are widely used in MIR \cite{bartsch2001catch},  and are based on the twelve-tone equal temperament system. Because notes that are exactly one octave apart sound very similar to the human ear, the distribution of chroma features, which ignores the specific octave, still captures important musical information. This approach can highlight perceived similarities between musical elements that might not be obvious in the original frequency spectrum. Chroma features are usually appointed as a 12-dimensional vector 
$v=[v(1), v(2), v(3),…, v(12)]$, where each component corresponds to one of the twelve pitch classes: $C, C\#, D, D\#, E, F, F\#, G, G\#, A, A\#$, and $B$. They show how the audio signal's energy is distributed among the twelve different pitch classes (notes in the chromatic scale).

\subsubsection{FilterBanks} 

A Mel filter bank consists of triangular filters specifically designed to approximate the way the human ear perceives differences in pitch and frequency. This design provides higher frequency resolution at lower frequencies and lower resolution at higher frequencies. Mel filter banks give more detail in low frequencies and less in high frequencies, which closely matches the way humans perceive sound \cite{fayek2016speech}.

\subsubsection{Mel-Spectrograms}

The signal is divided into frames, and a Fast Fourier Transform (FFT) is computed for each frame. Subsequently, a Mel-scale is applied, dividing the entire frequency spectrum into uniformly spaced bands. A spectrogram is then generated, where, for each frame, the signal magnitude is decomposed into its components corresponding to the frequencies in the Mel-scale.

\subsubsection{Mel-Frequency Cepstral Coefficients (MFCCs)}

For feature extraction in audio processing, MFCCs remain a standard choice across several sub-disciplines. Their utility is well-documented in the literature, ranging from the classification of music genres \cite{kizrak2014classification, thiruvengatanadhan2018music, mandal2020automatic, sharma2021classification} and the detection of emotional cues in music \cite{dutta2021music} to the broader requirements of speech recognition \cite{glittas2021low}. By mimicking the non-linear frequency perception of the human ear, MFCCs provide a psychoacoustically motivated framework for feature extraction, which remains a standard approach in speech recognition research. Their alignment with human auditory perception further extends their utility to music analysis, where capturing psychoacoustic nuances is essential \cite{li2011genre}. To capture the spectral characteristics of the signal, the MFCC pipeline first involves windowing the continuous waveform into discrete frames. A periodogram is then employed to estimate the power spectral density for each individual frame.

\subsection{Selected Classification Architecture Models}

Leveraging the robust feature-learning capabilities of CNN, prior studies have successfully deployed various architectures to categorize complex audio signals. In particular, AlexNet, VGG, and MobileNet have emerged as prominent choices for sound classification tasks. The following subsections offer a comparative overview of these architectures within the context of acoustic processing.

\begin{itemize}
	
	\item {\textbf{CNN}} architectural foundations of modern deep learning were established with the introduction of LeNet-5. While originally developed for character recognition, this pioneering framework introduced the essential concepts of local connectivity and shared weights through interleaved convolutional and subsampling layers. These principles proved vital for acoustic analysis, as they allow the network to achieve translation invariance, enabling the detection of specific sound patterns or pitch shifts regardless of their exact temporal position within a spectrogram.

	\item \textbf{AlexNet} \cite{krizhevsky2012imagenet} gained prominence following its decisive performance in the 2012 ImageNet Large Scale Visual Recognition Challenge (ILSVRC). This milestone is widely regarded as an inflection point for the field, as it demonstrated the potential of deep convolutional architectures to outperform traditional hand-crafted feature methods in complex pattern recognition tasks.
	
	\item \textbf{VGG} networks, developed by the Visual Geometry Group at Oxford in 2014, VGG-style architectures shifted the paradigm toward the systematic use of small $3 \times 3$ convolutional kernels. By stacking these filters in deep sequences, the design achieved a larger receptive field and increased model depth while simultaneously reducing the total number of parameters, a strategy that significantly enhanced its capacity for complex feature learning.

	
	\item \textbf{MobileNet}, Introduced by Howard et al. \cite{howard2017mobilenets}, MobileNet represents a specialized class of lightweight architectures engineered for deployment in resource-constrained environments. The core innovation lies in its use of depthwise separable convolutions, which decouple spatial filtering from feature generation. This architectural shift drastically minimizes both parameter count and computational overhead, facilitating real-time sound classification on embedded hardware without significant loss in accuracy.
	
\end{itemize}

\subsection{Proposed YMCM Architecture Model}

\begin{figure*}
	\centering
	\includegraphics[width=140mm]{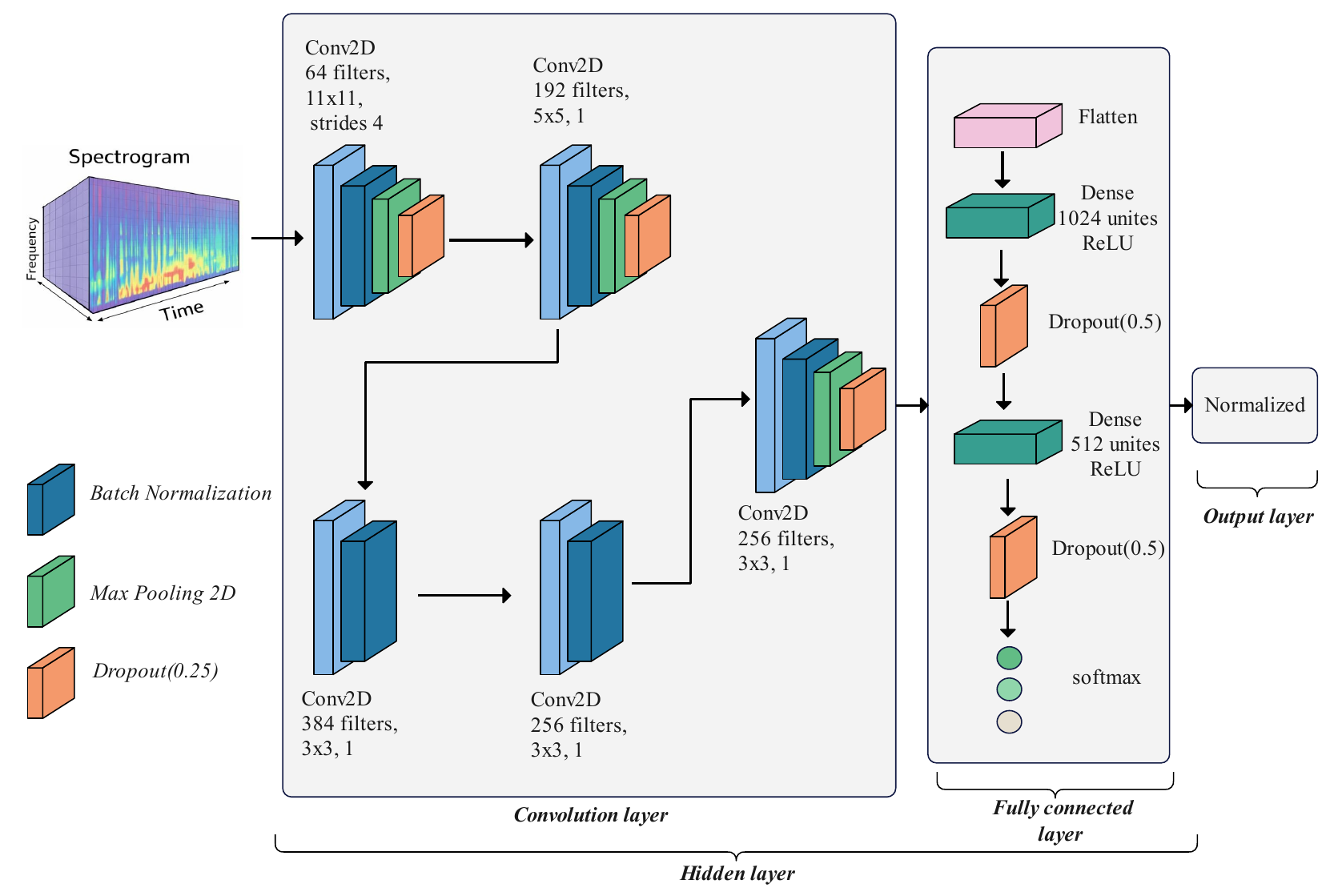}
	\caption{Workflow diagram of the proposed framework}
	\label{YMCM}
\end{figure*}

The adoption of CNNs in signal processing and music information retrieval is largely driven by their capacity for hierarchical representation learning. Within this framework, initial layers are typically sensitive to fundamental acoustic attributes, such as localized spectral textures and transient energy fluctuations. As the architecture deepens, these primitive features are aggregated into higher-order abstractions that correspond to complex musical properties, including timbre, harmonic density, and cepstral envelopes. This inherent ability to learn multi-scale features provides a robust motivation for employing CNN-based models in musical genre classification.

Drawing inspiration from the AlexNet framework, the proposed YMCM model utilizes a five-layer convolutional backbone with a filter distribution of 64, 192, 384, 256, and 256, respectively. The initial feature extraction is performed by an $11 \times 11$ kernel with a stride of 4, transitioning to a $5 \times 5$ kernel in the second layer, and $3 \times 3$ kernels for all subsequent stages. To ensure stable convergence and introduce non-linearity, each convolution is coupled with batch normalization and ReLU activation. Spatial dimensionality reduction is achieved via max-pooling layers using a $3 \times 3$ window. The architecture concludes with two dense layers—comprising 1024 and 512 neurons—leading to a final softmax-activated layer for genre categorization, as depicted in Fig.~\ref{YMCM}.

\section{Experiments and results }

This section details the datasets, training configurations, and performance metrics utilized to validate the proposed framework. The experimental design is structured to provide a comprehensive evaluation of the system's robustness and generalizability across diverse acoustic environments
\subsection{Training Protocol}

In this study, standard implementations of CNN, AlexNet, VGG16, and MobileNet were utilized with their original network architectures and default parameter configurations. In contrast, the Yemeni Music Classification Model (YMCM) was implemented as a CNN, with architectural modifications.

Audio feature extraction was performed using well-established Python-based libraries. Chroma, Mel Spectrogram, and MFCC features were extracted using the \textit{librosa} library (version 0.11.0) \cite{mcfee2015librosa}, while FilterBank features were obtained using the \textit{Python Speech Features} library (version 0.6). We derived Mel-spectrograms utilizing 128 discrete frequency bands, Chroma features with 12 pitch classes, and MFCC features with 13, 20, and 40 coefficients, following the standard configurations provided by the respective libraries.

Each classification model was trained using a single feature type per experimental configuration. For the experiments reported in this work, MFCC features with 13, 20, and 40 coefficients were adopted. The models were implemented using the Keras deep learning framework (version 2.9.0) with a TensorFlow 2.9.0 backend. All experiments were conducted on a workstation running Python 3.8 on Ubuntu 20.04, equipped with one virtual GPU (vGPU) with 32 GB of memory. GPU acceleration was enabled using the NVIDIA CUDA Toolkit version 11.2 to enhance computational efficiency during training.

The network was trained using the Adam optimizer with a fixed learning rate of $10^{-4}$. To minimize the discrepancy between predicted and ground-truth genre labels, we employed categorical cross-entropy as the objective function. The training process was capped at 50 epochs with a mini-batch size of 16; however, to prevent overfitting and ensure optimal generalization, an early stopping mechanism was implemented. This strategy monitored the validation loss and terminated training if no improvement was observed for 10 consecutive epochs. The specific hyperparameter configurations are consolidated in Table \ref{tab:training_parameters}.

\begin{table}[h]
	\centering
	\caption{Training Parameters Used in the Experiments}
	\label{tab:training_parameters}
	\begin{tabular}{ll}
		\hline
		\textbf{Parameter} & \textbf{Value} \\
		\hline
		Number of Epochs & 50 \\
		Batch Size & 16 \\
		Learning Rate & 0.0001 \\
		Optimizer & Adam \\
		Loss Function & Categorical Crossentropy \\
		Early Stopping Patience & 10 epochs \\
		\hline
	\end{tabular}
\end{table}

\subsection{Evaluation Metrics}

The performance of the proposed Yemeni Music Classification Model (YMCM) system is evaluated using several key metrics: weighted precision, sensitivity (Recall), F1-score, specificity, and balanced accuracy. These metrics, which assess various aspects of classification performance, were calculated as shown in the following formulas:

\begin{equation}
	\text{Precision} = \frac{TP}{TP + FP}
	\tag{1}
\end{equation}

\begin{equation}
	\text{Recall} = \frac{TP}{TP + FN}
	\tag{2}
\end{equation}

\begin{equation}
	\text{F1-score} = \frac{2 \times \text{Precision} \times \text{Recall}}{\text{Precision} + \text{Recall}}
	\tag{3}
\end{equation}

\begin{equation}
	\text{Specificity} = \frac{TN}{FP + TN}
	\tag{4}
\end{equation}

\begin{equation}
	\text{Accuracy} = \frac{1}{n} \sum \frac{TP}{FP + TN + TP + FN}
	\tag{5}
\end{equation}

\begin{figure*}
	\centering
	\includegraphics[width=140mm]{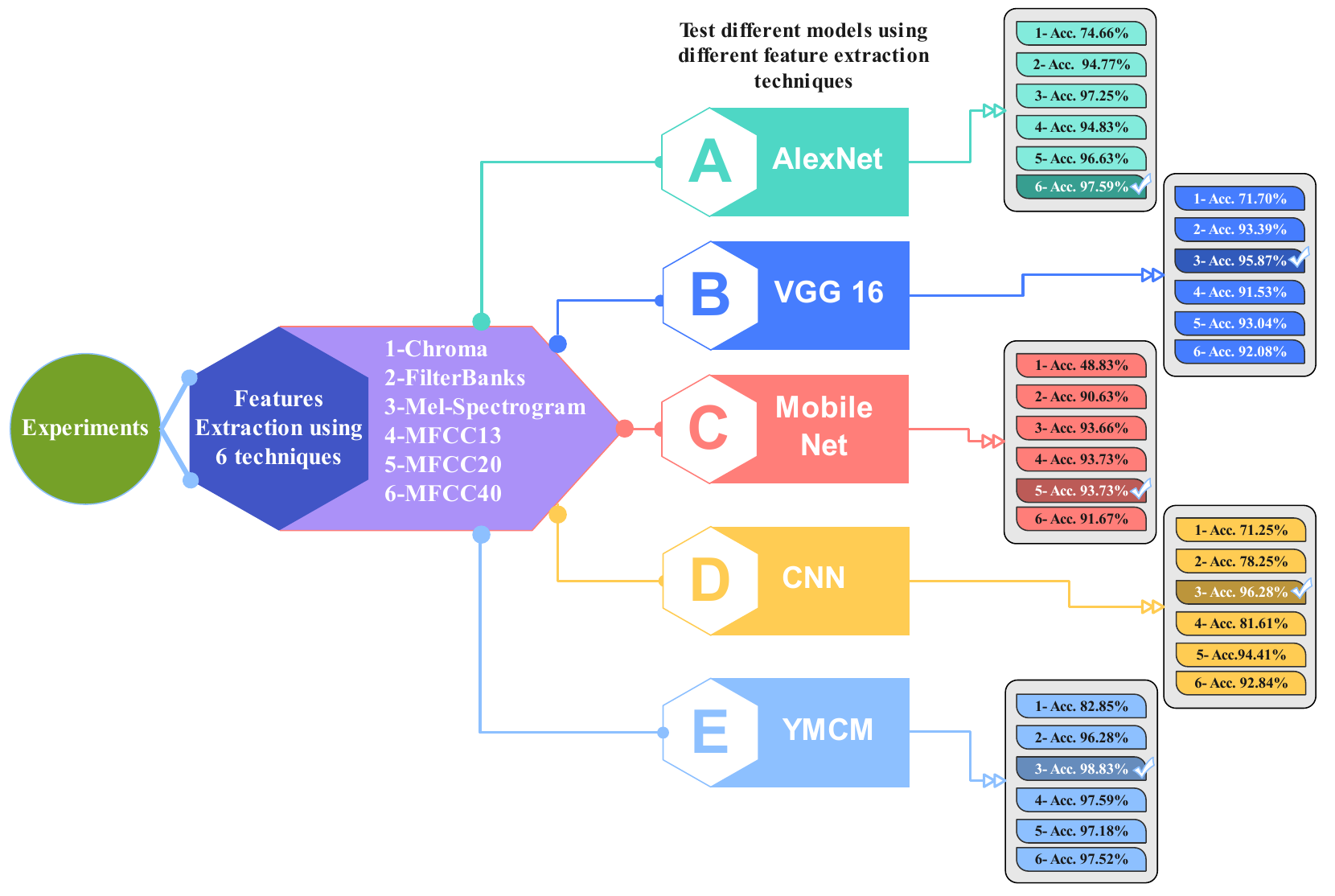}
	\caption{Summary of the experimental design and configurations conducted in this study.}
	\label{experiments}
\end{figure*}

The performance of the model is quantified using the fundamental components of the confusion matrix: True Positives (TP), False Positives (FP), True Negatives (TN), and False Negatives (FN). In this context, TP and TN represent instances where the predicted labels align correctly with the ground truth for positive and negative classes, respectively. Conversely, FP and FN denote misclassifications, where the model erroneously assigns a positive label to a negative instance or fails to identify a true positive case.

\subsection{Experiment Steps}

As illustrated in Fig.~\ref{experiments}, the experimental evaluation is organized into six main experimental groups, each comprising five sub-experiments. In the six main experiments, different time–frequency feature extraction techniques are employed, including MFCCs with 13, 20, and 40 coefficients (MFCC13, MFCC20, and MFCC40), Mel spectrograms, Chroma features, and FilterBank features, which are extracted from the YMIR dataset. For each feature representation, the extracted features are independently used as input to five CNN models, namely the proposed YMCM, AlexNet, CNN, VGG16, and MobileNet. This experimental configuration results in a total of 30 distinct experiments.

\subsection{Discussion and results}

An evaluation of the YMCM framework across various feature extraction methodologies reveals a high degree of robustness, with the model maintaining consistently strong classification accuracy, as shown in Table~\ref{tab:performance_comparison}. When Chroma features are used, the model achieves an accuracy of 82.85\%, which reflects the limited discriminative capacity of pitch-class information that omits spectral energy distribution and temporal detail. The use of FilterBank features substantially improves performance to 96.28\% accuracy by capturing perceptually motivated spectral energies; however, these representations do not explicitly preserve fine-grained time–frequency continuity. The highest performance is achieved with Mel-Spectrogram features, where YMCM attains up to 98.83\% accuracy, primarily because Mel-Spectrograms preserve detailed spectral-temporal structures over time, enabling the model to jointly learn short-term dynamics, such as onsets and transients, as well as longer-term patterns related to rhythm and timbre. In contrast, MFCC-based representations achieve accuracies of 97.59\%, 97.18\%, and 97.52\% for MFCC13, MFCC20, and MFCC40, respectively; however, their reliance on a discrete cosine transform compresses the spectral representation and partially removes frequency locality, which limits the ability of convolutional layers to exploit spatial correlations. Overall, these results demonstrate that while YMCM effectively leverages both compact cepstral and dense spectral representations, Mel-Spectrograms provide the most informative and discriminative features due to their preservation of spectral continuity and strong compatibility with convolutional feature learning.

\begin{figure}
	\centering
	\includegraphics[width=85mm]{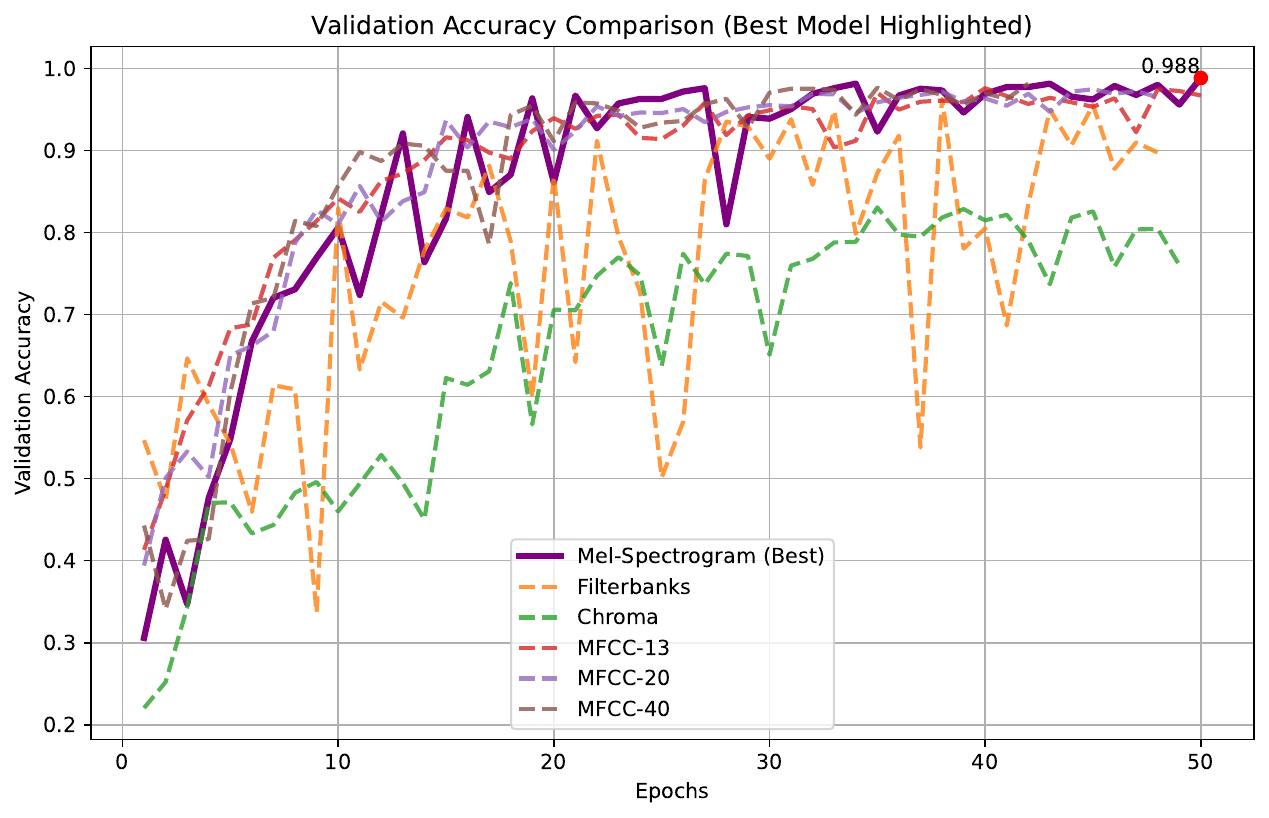}
	\caption{Validation accuracy of the proposed YMCM framework using several feature extraction methods.}
	\label{fig:validation_accuracy}
\end{figure}

\begin{figure}
	\centering
	\includegraphics[width=85mm]{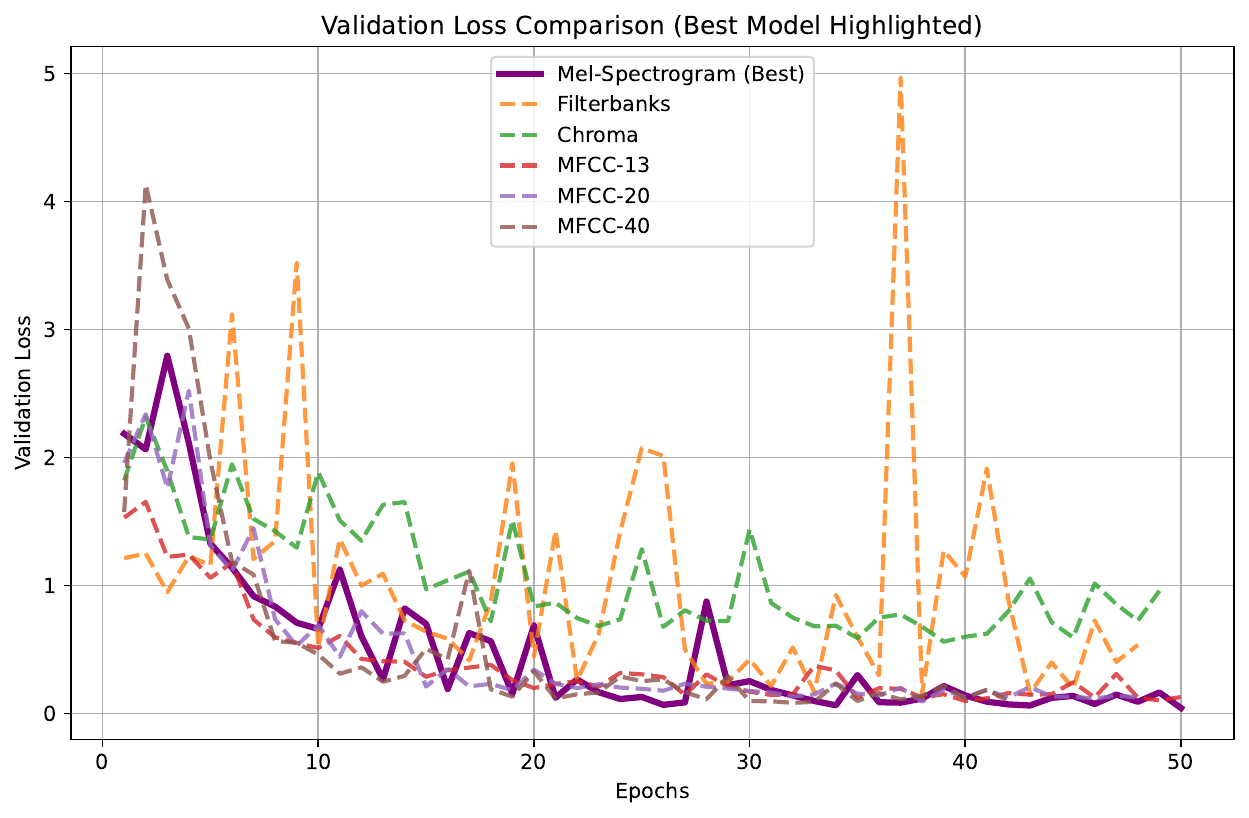}
	\caption{Validation loss of the proposed YMCM framework using several feature extraction methods.}
	\label{fig:validation_loss}
\end{figure}

The validation loss and accuracy curves in Figs.~\ref{fig:validation_accuracy} and ~\ref{fig:validation_loss} illustrate the training behavior of he YMCM model with different feature extraction techniques. Among all features, the Mel-Spectrogram consistently exhibits the most stable convergence, characterized by a rapid reduction in validation loss and a smooth increase in validation accuracy, ultimately achieving the highest accuracy of approximately 0.99. In contrast, FilterBanks and Chroma features show noticeable fluctuations in both loss and accuracy, indicating less stable optimization and weaker generalization. MFCC-based features (MFCC13, MFCC20, and MFCC40) demonstrate improved convergence compared to Chroma and FilterBanks; however, they still exhibit slightly higher variance and slower stabilization than Mel-Spectrograms.

\begin{figure*}[!t]
	\centering
	\subfloat[Filterbanks]{\includegraphics[width=0.3\linewidth]{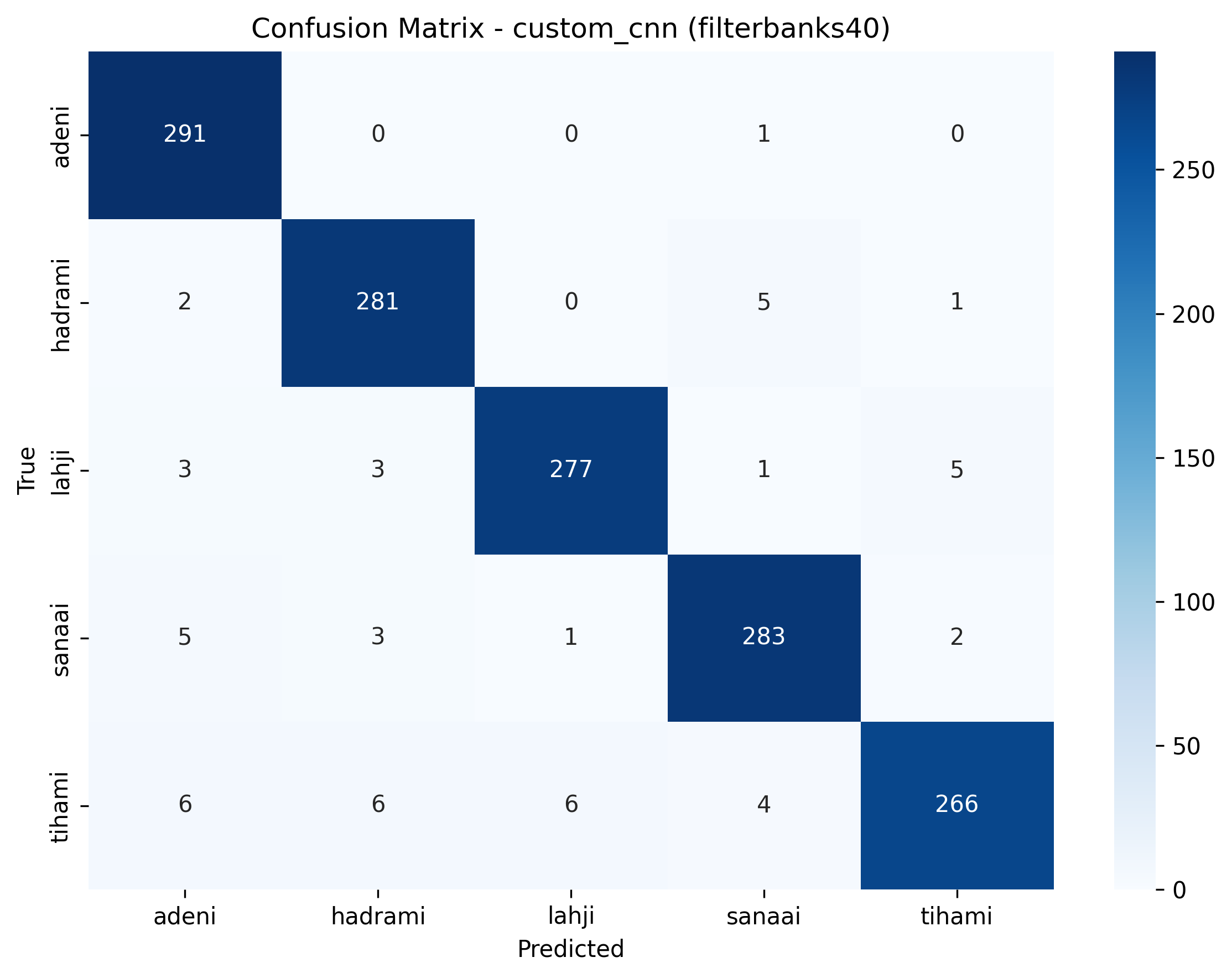}%
		\label{fig:cnn_filterbanks40}}
	\hfil
	\subfloat[Chroma]{\includegraphics[width=0.3\linewidth]{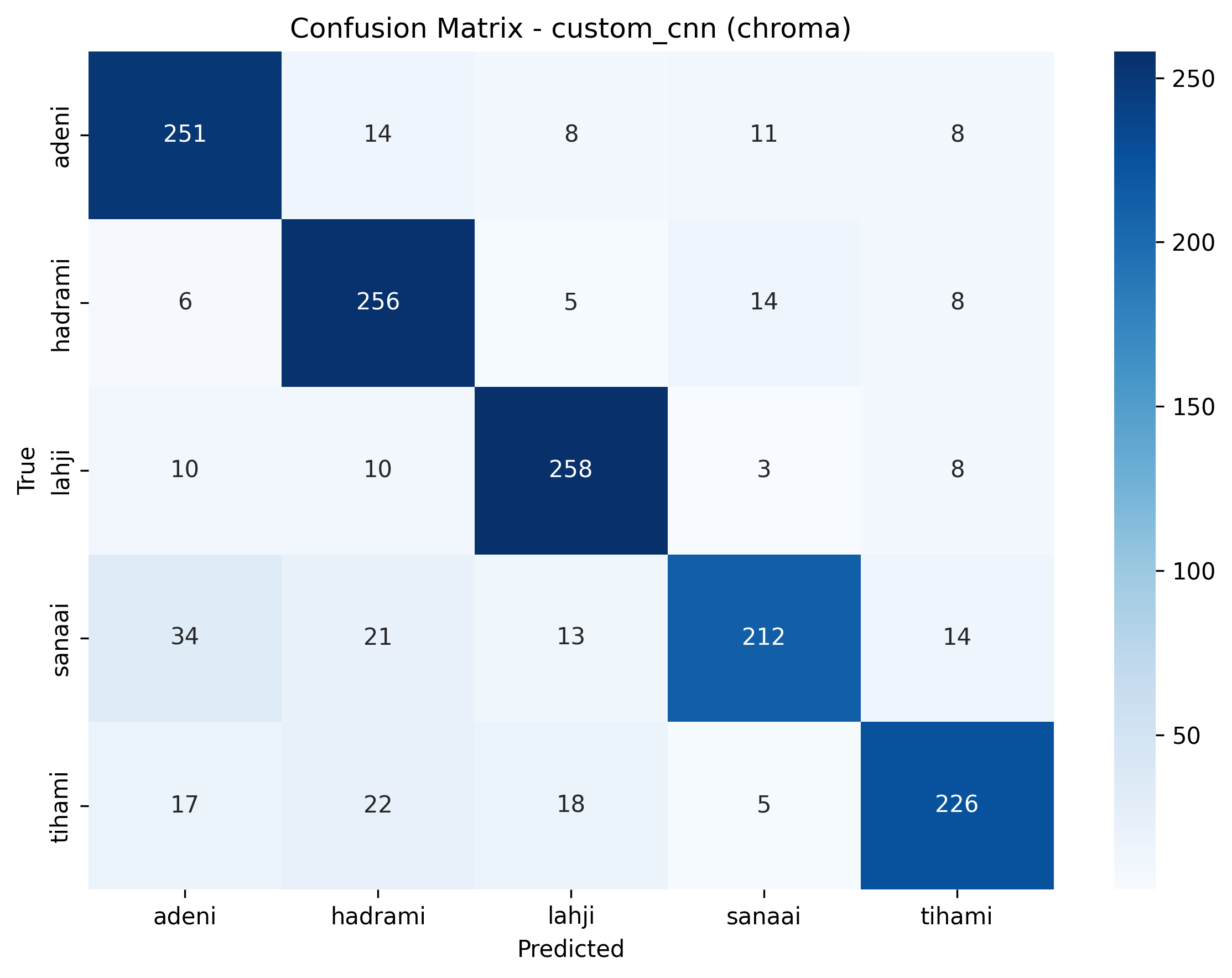}%
		\label{fig:cnn_chroma}}
	\hfil
	\subfloat[Mel Spectrogram]{\includegraphics[width=0.3\linewidth]{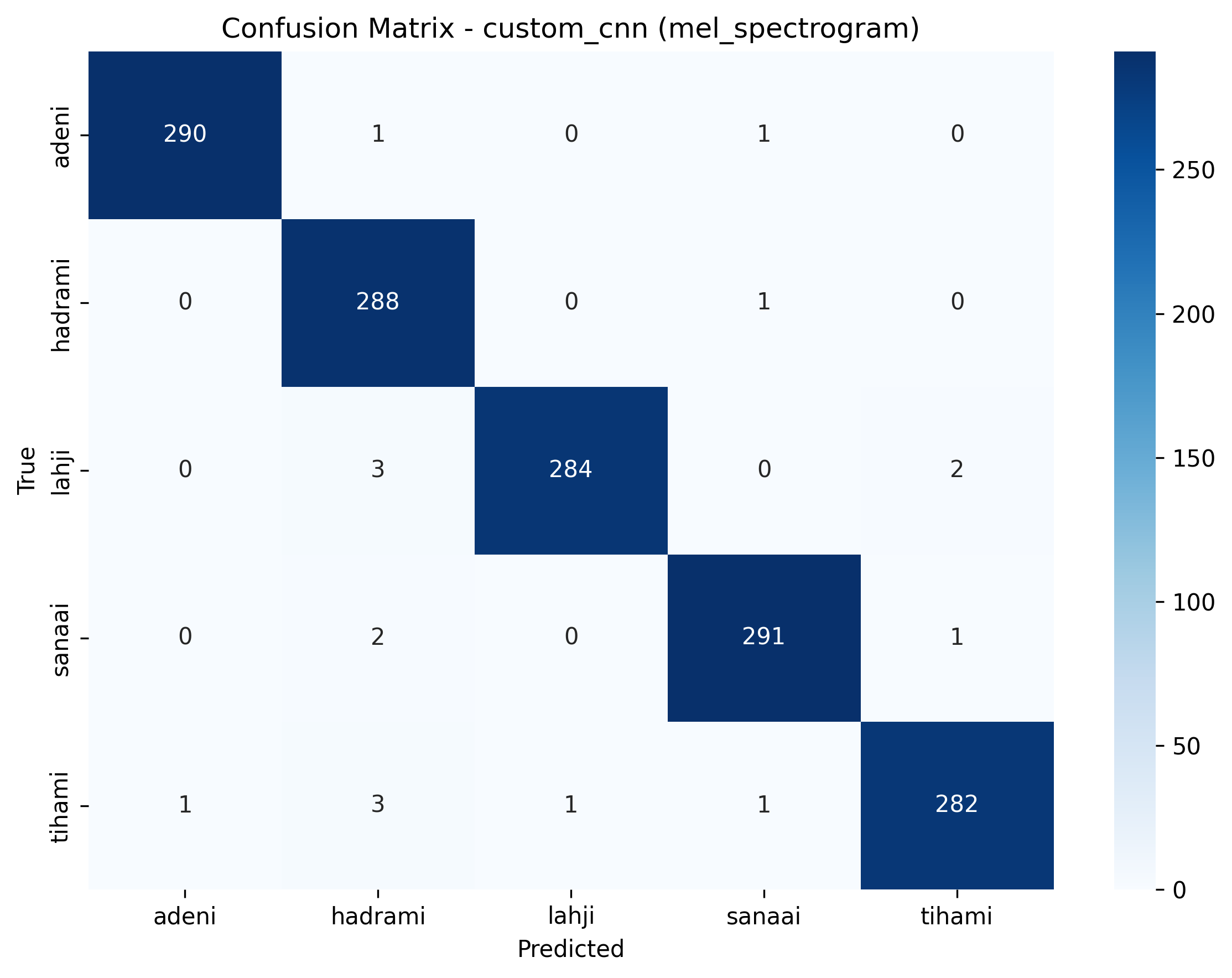}%
		\label{fig:cnn_mel_spectrogram}}

	\vspace{0.3cm}
	
	\subfloat[MFCC 13]{\includegraphics[width=0.3\linewidth]{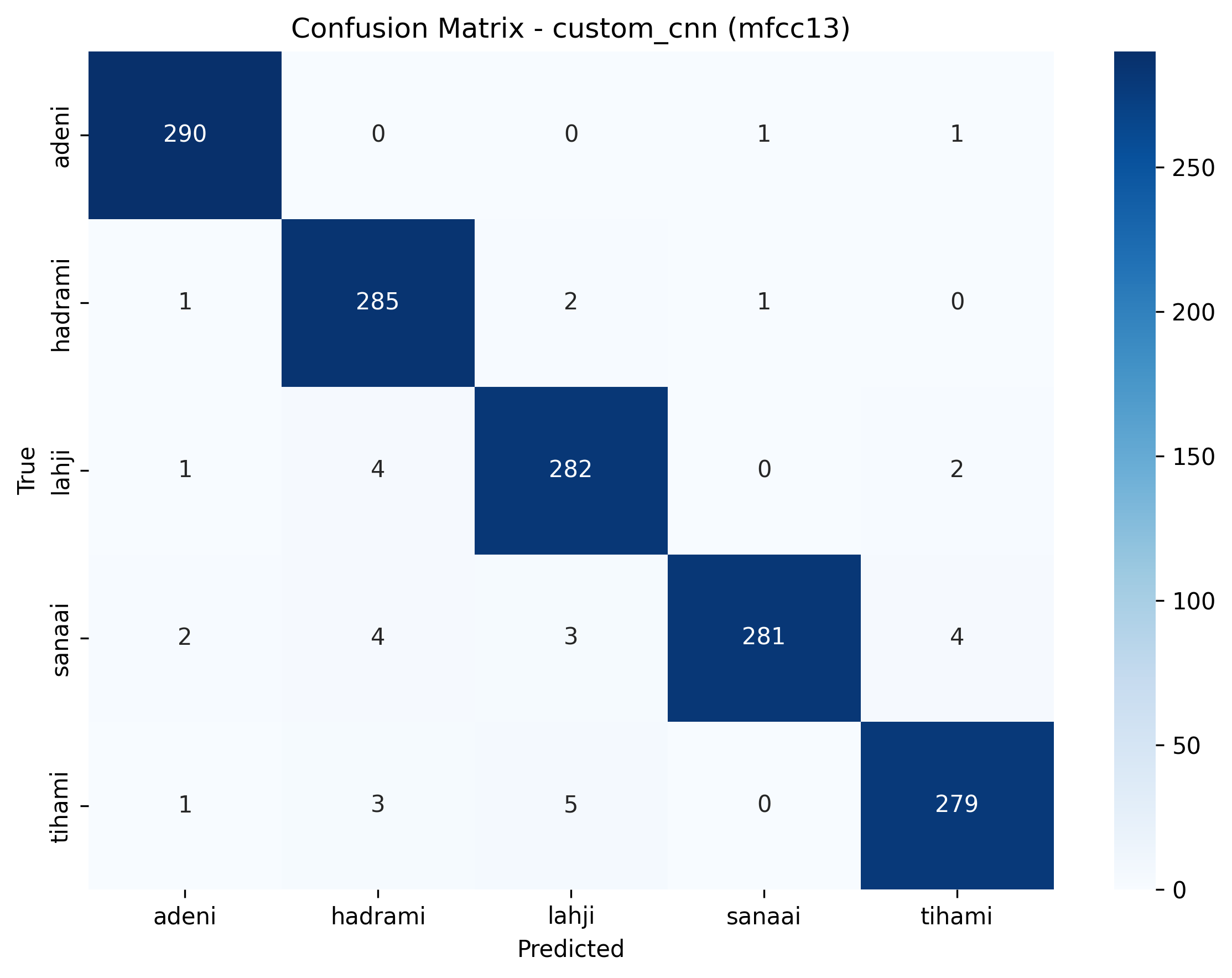}%
		\label{fig:cnn_mfcc13}}
	\hfil
	\subfloat[MFCC 40]{\includegraphics[width=0.3\linewidth]{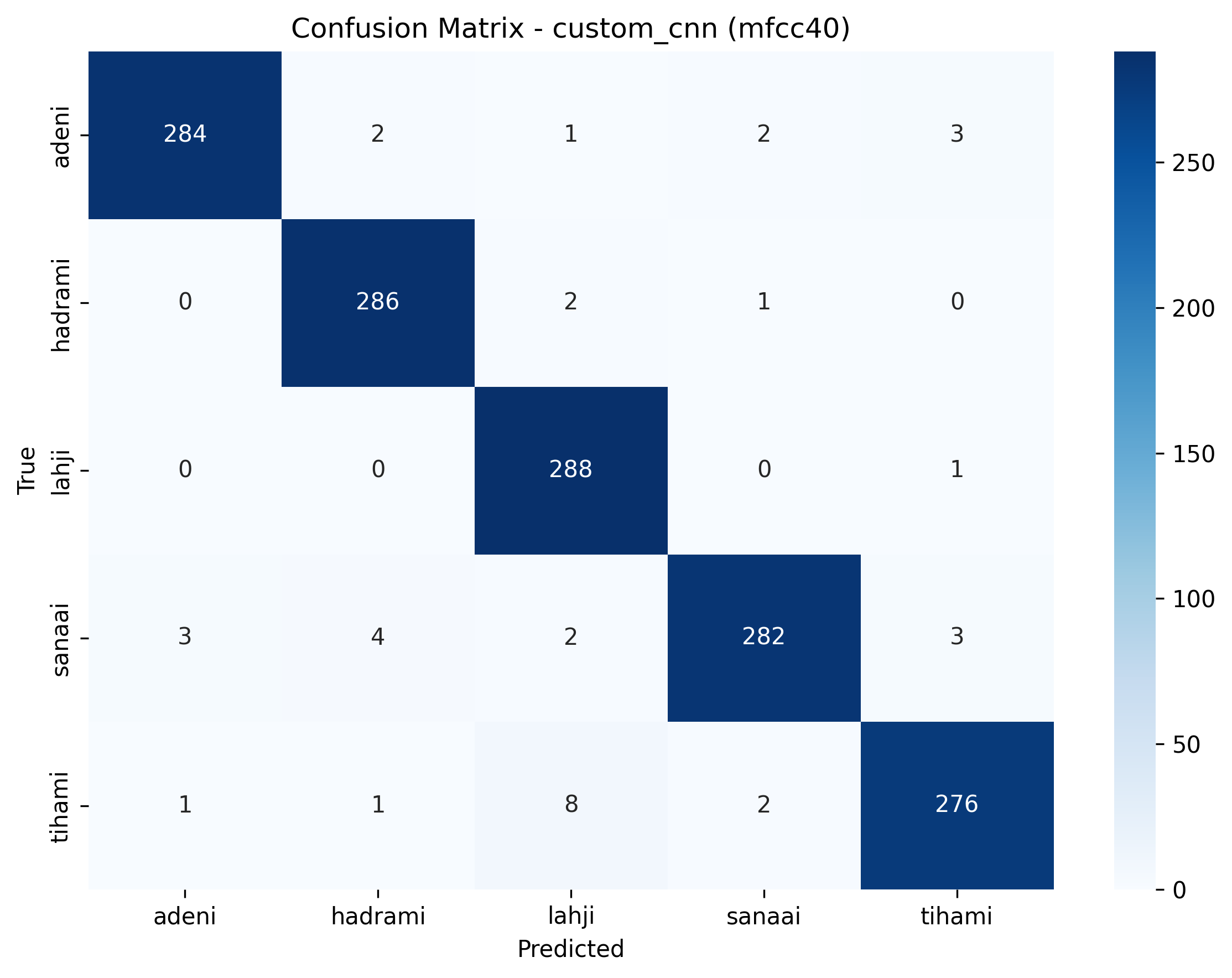}%
		\label{fig:cnn_mfcc40}}
	\hfil
	\subfloat[MFCC 20]{\includegraphics[width=0.3\linewidth]{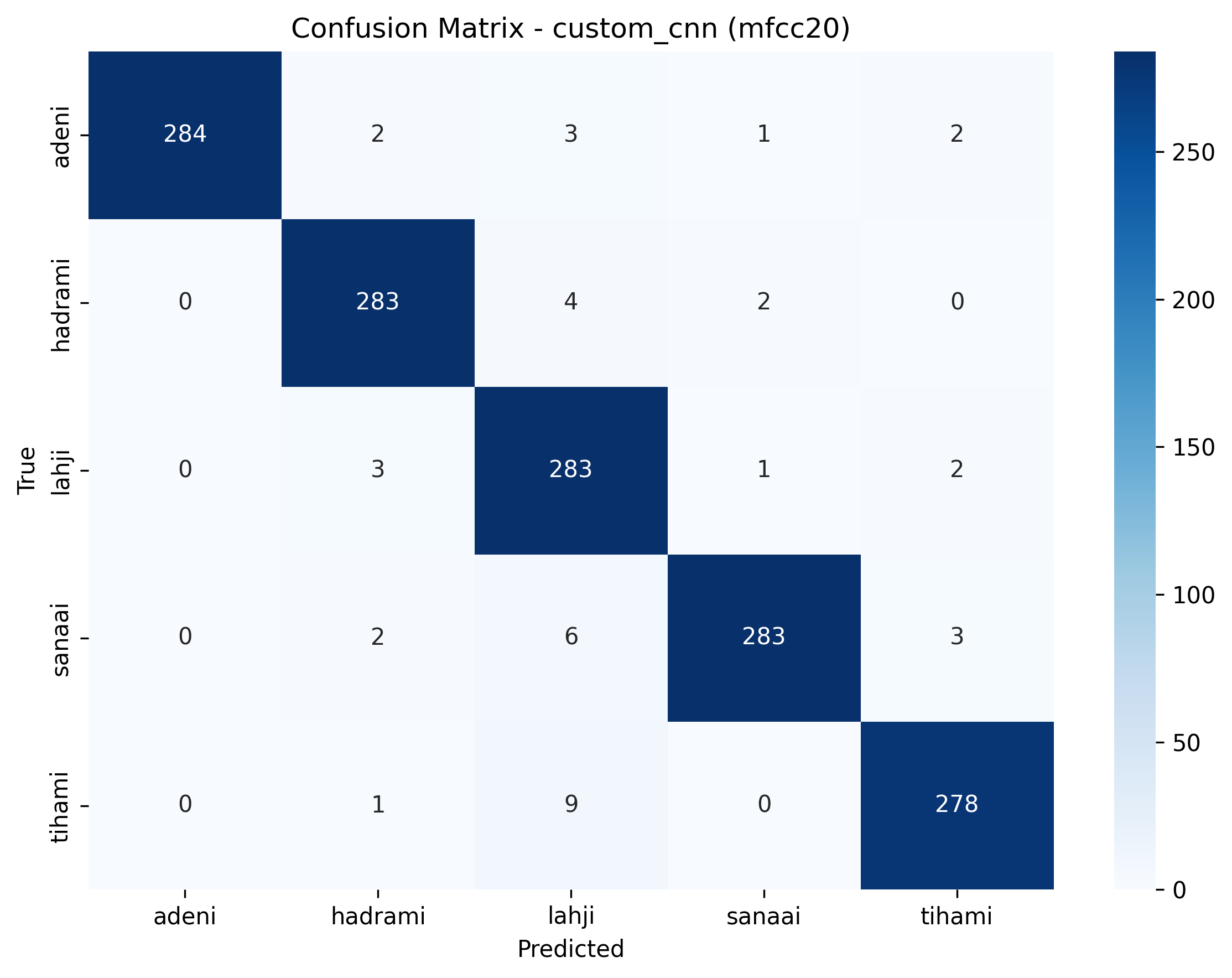}%
		\label{fig:cnn_mfcc20}}
	
	\caption{Confusion matrices for the proposed YMCM model with different audio feature extraction methods: (a) Filterbanks, (b) chroma features, (c) mel spectrogram, (d) 13 MFCCs, (e) 40 MFCCs, (f) 20 MFCCs. All matrices represent a 5-class classification task.}
	\label{fig:cnn_confusion_matrices}
\end{figure*}

The confusion matrices for each feature type, processed through the YMCM model, are presented in Fig.~\ref{fig:cnn_confusion_matrices}. The mel spectrogram features achieved the highest classification accuracy, with strong diagonal dominance (290–291 correct predictions per class) and minimal misclassification. Filterbanks and MFCC variants also performed competitively, though with slightly increased inter-class confusion. In contrast, chroma features exhibited the weakest performance, with significant off-diagonal errors particularly for Class 4 and Class 5, where only 212 and 226 samples were correctly classified, respectively.

\begin{table}
	\centering
	\caption{Performance Results of Different Models Using Various Feature Extraction Techniques.}
	\label{tab:performance_comparison}
	\renewcommand{\arraystretch}{1.2}
	\begin{tabular}{llcccc}
		\hline
		\textbf{Model} & \textbf{Feature } & \textbf{Acc.} & \textbf{Prec.} & \textbf{Rec.} & \textbf{F1.} \\
		& \textbf{Extraction} & \textbf{(\%)} & \textbf{(\%)} & \textbf{(\%)} & \textbf{(\%)}
		\\
		\hline
		
		\multirow{6}{*}{AlexNet}
		& Chroma           & 74.66 & 75.20 & 74.66 & 74.64 \\
		& FilterBanks      & 94.77 & 94.79 & 94.77 & 94.76 \\
		& Mel-Spectrogram  & 97.25 & 97.27 & 97.25 & 97.24 \\
		& MFCC13           & 94.83 & 95.03 & 94.83 & 94.84 \\
		& MFCC20           & 96.63 & 96.76 & 96.63 & 96.64 \\
		& MFCC40           & 97.59 & 97.61 & 97.59 & 97.59 \\
		\hline
		
		\multirow{6}{*}{VGG16}
		& Chroma           & 71.07 & 70.87 & 71.07 & 70.73 \\
		& FilterBanks      & 93.39 & 93.40 & 93.39 & 93.39 \\
		& Mel-Spectrogram  & 95.87 & 95.90 & 95.87 & 95.87 \\
		& MFCC13           & 91.53 & 91.57 & 91.53 & 91.52 \\
		& MFCC20           & 93.04 & 93.20 & 93.04 & 93.06 \\
		& MFCC40           & 92.08 & 92.12 & 92.08 & 92.07 \\
		\hline

		& Chroma           & 48.83 & 47.67 & 48.83 & 47.49 \\
		& FilterBanks      & 90.63 & 90.69 & 90.63 & 90.64 \\
		Mobile& Mel-Spectrogram  & 93.66 & 93.70 & 93.66 & 93.65 \\
		Net& MFCC13           & 93.73 & 93.75 & 93.73 & 93.73 \\
		& MFCC20           & 93.73 & 93.75 & 93.73 & 93.72 \\
		& MFCC40           & 91.67 & 91.73 & 91.67 & 91.67 \\
		\hline
		
		\multirow{6}{*}{CNN}
		& Chroma           & 71.25 & 62.02 & 70.25 & 65.02 \\
		& FilterBanks      & 78.25 & 78.89 & 68.25 & 68.58 \\
		& Mel-Spectrogram  & 96.28 & 96.53 & 96.28 & 96.32 \\
		& MFCC13           & 81.61 & 85.67 & 81.61 & 81.89 \\
		& MFCC20           & 94.35 & 94.41 & 94.35 & 94.35 \\
		& MFCC40           & 92.84 & 93.47 & 92.84 & 92.90 \\
		\hline

		& Chroma           & 82.85 & 83.15 & 82.85 & 82.75 \\
		& FilterBanks      & 96.28 & 96.31 & 96.28 & 96.27 \\
		Proposed& Mel-Spectrogram  & 98.83 & 98.84 & 98.83 & 98.83 \\
		YMCM& MFCC13           & 97.59 & 97.61 & 97.59 & 97.59 \\
		& MFCC20           & 97.18 & 97.25 & 97.18 & 97.19 \\
		& MFCC40           & 97.52 & 97.54 & 97.52 & 97.52 \\
		\hline
		
	\end{tabular}
\end{table}

\begin{table}[b!]
	\centering
	\caption{Optimal Feature Extraction Strategy for Each Model}
	\label{tab:optimal_features}
	\renewcommand{\arraystretch}{1.2}
	\begin{tabular}{llc}
		\hline
		\textbf{Model} & \textbf{Optimal Feature} & \textbf{Accuracy} \\
		\hline
		AlexNet        & MFCC40           & 97.59\% \\
		VGG16          & Mel-Spectrogram  & 95.87\% \\
		MobileNet      & MFCC13 / MFCC20  & 93.73\% \\
		CNN (Baseline) & Mel-Spectrogram  & 96.28\% \\
		YMCM (Proposed)& Mel-Spectrogram  & 98.83\% \\
		\hline
	\end{tabular}
\end{table}

Based on the results reported in Table \ref{tab:optimal_features},  the optimal feature extraction strategy depends on the underlying model architecture. For AlexNet, MFCC40 achieves the highest accuracy 97.59\%, indicating that higher-dimensional cepstral representations effectively complement its large receptive fields. VGG16 and the baseline CNN attain their best performance using Mel-Spectrogram features, 95.87\% and 96.28\%, respectively, highlighting the importance of preserving detailed time–frequency structures for deeper convolutional networks. For the lightweight MobileNet architecture, MFCC13/MFCC20 provides the best trade-off between accuracy 93.73\% and computational efficiency. The proposed YMCM model achieves its highest accuracy with Mel-Spectrograms at 98.83\%, confirming that dense spectral-temporal representations are particularly well suited to its design. Overall, these results indicate that Mel-Spectrograms are more effective for deeper and more expressive convolutional architectures, whereas MFCC-based features offer a more efficient alternative for lightweight models. Eventually, under identical experimental conditions, the YMCM model consistently outperforms all benchmark architectures listed. These findings suggest that the YMCM model more effectively leverages complex audio representations, demonstrating a superior capacity for generalization. The consistency of these results across varied acoustic conditions underscores the model’s robustness for high-fidelity musical genre categorization.


\section{Conclusion}

This paper introduced a comprehensive end-to-end framework specifically engineered for the classification of Yemeni musical genres, addressing the significant underrepresentation of culturally distinct Arabic traditions in the current Music Information Retrieval (MIR) landscape. Central to this contribution is the YMIR dataset, an expert-annotated corpus encompassing five foundational genres, validated through rigorous inter-annotator agreement to ensure high label fidelity. By deploying the YMCM architecture, we demonstrate that a CNN-based approach tailored for time-frequency representations can effectively capture the unique rhythmic and melodic nuances of Yemeni music within a standardized experimental framework. Extensive experiments across multiple feature representations demonstrated that dense spectral–temporal features are particularly effective for deep convolutional learning in this task. The proposed YMCM achieved the best overall performance, reaching 98.83\% accuracy with Mel-spectrogram features and outperforming AlexNet, VGG16, MobileNet, and a baseline CNN under the same conditions. The results further indicate that feature selection is closely coupled with model capacity: Mel-spectrograms typically benefit deeper, more expressive CNNs, whereas MFCC variants remain competitive for lightweight architectures. 

Future work will extend YMIR in both scale and diversity (e.g., broader regional coverage and additional sub-genres) and investigate more advanced paradigms such as transformer-based audio encoders, self-supervised representation learning, and cross-dataset transfer learning to improve robustness in low-resource music classification settings.

\section*{AI Statement}
We used AI tools to improve grammar, readability, and wording. We did not use AI to generate data, analyses, figures, or conclusions. The authors reviewed and verified all content and take full responsibility for it.

\section*{Acknowledgement}
This work was supported by Yidan University Education Foundation under Grant JJA202507.

\bibliographystyle{IEEEtran}    
\bibliography{IEEE_TAES_regular_template_latex}

\vfill

\end{document}